\documentclass[pre,twocolumn,superscriptaddress,showpacs]{revtex4}
\usepackage{graphicx}
\usepackage{amsmath}

\begin{document}

\title{Knot Probability for Self-Avoiding Loops on a Cubic Lattice}
\author{Yacov Kantor}
\email{kantor@post.tau.ac.il}
\affiliation{School of Physics and
Astronomy, Tel Aviv University, Tel Aviv 69978, Israel}
\author{Mehran Kardar}
\affiliation{Department of Physics, Massachusetts
Institute of Technology, Cambridge,
Massachusetts 02139, USA}

\date{\today}

\begin{abstract}
We investigate the probability for appearance of knots in
self--avoiding loops (SALs) on a cubic lattice. A set of $N$--step
loops is generated by attempting to combine pairs of $N/2$--step
self--avoiding walks constructed by a dimerization method. We
demonstrate that our method produces unbiased samples of SALs, and
study the knot formation probability as a function of loop size.
Our results corroborate the conclusions of {\em Yao et. al.} with
loops generated by a Monte Carlo method\cite{yao}.
\end{abstract}

 \maketitle

Knots and links naturally appear in long polymers
\cite{catenanes}, and play a prominent role in biological systems
and processes\cite{frisch}. Examples include chromosomes during
cell division \cite{alberts}, knots in bacterial DNA
\cite{ringclosure}, or knots in the native states of
proteins\cite{taylor}. It can be shown rigorously that very long
self--avoiding loops are always knotted\cite{sumners}. However,
the theoretical proofs do not provide the frequency of the knots,
or the functional dependence of their frequency on loop size.
Quantitative insight into this question was first provided by a
numerical study of random walks on a lattice \cite{volog}.
Excluded volume effects,  or self--avoiding (SA) interactions, are
certainly crucial for correct description of
polymers\cite{degennesSC}; however, their incorporation into
numerical studies is not simple. Earlier studies considered
continuum models of self--avoiding loop (SAL) polymers with
varying degrees of self--repulsion \cite{koniar,deguchi}, and
demonstrated that with increasing number of monomers $N$, the
fraction of unknotted loops decreases as ${\rm e}^{-N/N_o}$. The
characteristic size at which knots appear is surprisingly large:
It increases from several hundred steps in the absence of
self--avoidance, to hundreds of thousands for strongly SA
polymers. Since the value of $N$ used in typical simulations does
not exceed several thousands, for SAL one can assume that the
probability of an unknotted configuration simply decays
exponentially. The value of  $N_o$ can then be extracted by noting
that for $N\ll N_o$ the probability of the knotted configurations
is $P_N\approx N/N_o$. A recent study by {\em Yao et
al.}\cite{yao} investigated the knotting probability of SALs on a
cubic lattice with $N\le 3000$ and found $N_o\approx 2.5\times
10^5$. Our results corroborate the above study using a different
approach to generating SALs.

\begin{figure}
\includegraphics[width=8cm]{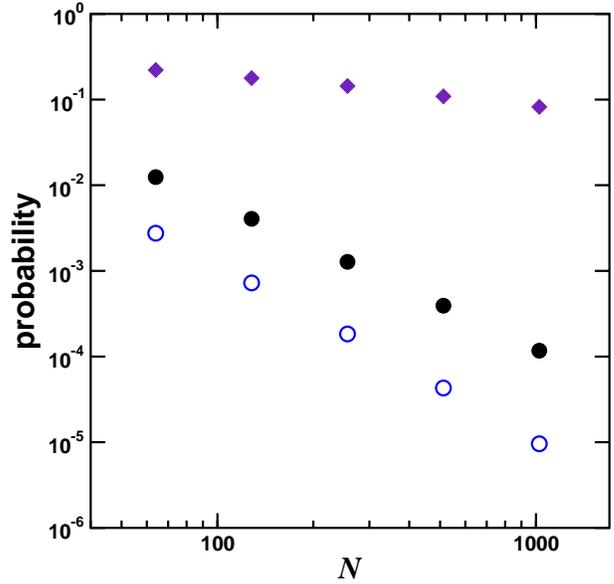}
\caption{\label{loop_prob} The probability that two $N/2$--step
SAWs with the same origin terminate at the same point (full
circles), and the probability of their forming a proper SAL (open
circles). (These results were obtained in our simulations by
including a 48--fold symmetry factor enhancement, as explained in
the text.) The ratio between the latter and the former, i.e. the
probability that the loop formed by the $N/2$-step pairs is
self--avoiding (diamonds) as a function of $N$.}
\end{figure}

Generating sufficiently large numbers of SALs has been the main
obstacle to the study of statistics of knots in polymers. Static
methods for creating SA polymers one at a time have the advantage
of producing configurations that are independent of each other. If
we do not require the two ends of a polymer to meet, several
methods generate samples of ``properly weighted" configurations:
For lattice or for continuum models with ``hard" potentials, this
means that every configuration has the same weight. In the
dimerization algorithm\cite{dimerization}, an $N$-step SA walk
(SAW) is created by generating two $(N/2)$-step SAWs and
attempting to concatenate them. (If the concatenated walk is
self--intersecting, it is discarded and the pair creation process
is restarted.) The resulting SAWs are properly weighted, i.e. each
has the same weight. While this method is very
efficient\cite{madrassokal}, it is not well suited for generating
SALs: To produce a properly weighted $2N$--step SAL, we could
first generate two $N$--step SAWs, assume that they both start at
the origin, and check whether they do not intersect and end at the
same point creating a loop, {\em discarding the pair if they do
not form a proper SAL}. However, the probability of two SAWs in
$d$ dimensions accidentally terminating at the same point is
proportional to $R^{-d}\sim N^{-d\nu}$, where $R^2$ is the mean
squared end-to-end distance of $N$-step SAWs, and in $d=3$ the
swelling exponent is $\nu=0.588$ \cite{madrasslade}. Full circles
in the Fig.~\ref{loop_prob} depict  this decay with $N$. In
addition, the probability that two such walks do not intersect
each other decays as $N^{2(1-\gamma)}$, where $\gamma=1.158$ in
$d=3$\cite{cara}, as demonstrated by the diamonds in
Fig.~\ref{loop_prob}. Thus, successful formation of a
three-dimensional SAL using this {\em direct} method is
proportional to $N^{-2.08}$, as depicted by open circles in
Fig.~\ref{loop_prob}.

\begin{figure}
\includegraphics[width=8cm]{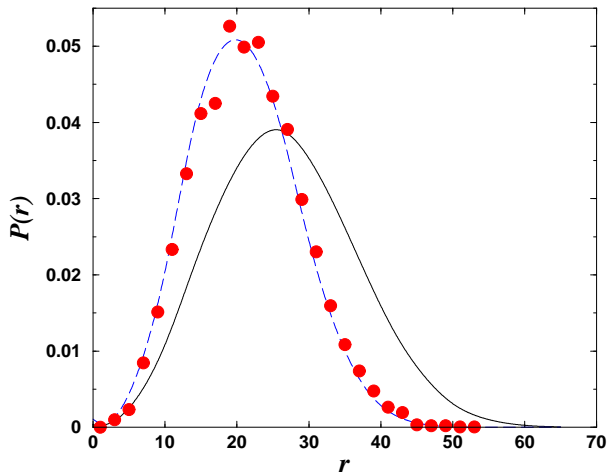}
\caption{\label{compar_prob} The solid line depicts the
probability distribution for the end to end distance of a single
self-avoiding walk of length 256. The dashed line corresponds to
the ``correct ensemble'' in which two such walks accidentally meet
at the same position. In Ref.\protect{\cite{chen}}, the former
ensemble is used to generate instances of the latter, weighting
the resulting configurations with the appropriate bias. The full
circles are the distribution of the end-to-midpoint distance of
actual, successfully generated, 512--step SALs in our dimerization
procedure.}
\end{figure}

To circumvent the problem of the rarity of cases where two SAWs
starting from the origin terminate at the same point, Chen
suggested \cite{chen} accumulating  lists of SAWs. Whenever two
SAWs in the list have the same end-to-end distance, they are
rotated to make their endpoints coincide, thus producing a loop.
Keeping the information on large numbers of SAWs is computationally
memory--intensive. A more serious concern is that the ensemble of
generated loops is {\em biased}: The solid line in
Fig.~\ref{compar_prob} depicts the probability distribution of the
end-to-end distance $r$ for 256-step SAWs; the dashed line
corresponds to the subset of cases where a pair of SAWs has the
same end--point. The latter is the correct ensemble for loops,
while the former is the ensemble produced when accumulated SAWs
are rotated and linked. This distinction in weight (bias) is
well-known, and is properly corrected in Chen's algorithm.
However, as in all bias-corrected methods, while the expectation
value of a desired quantity is correctly reproduced, the variance
of this quantity increases with system size. Essentially,
bias--corrected methods attempt to reconstruct one distribution
from a tail of another distribution; the distributions moving
further apart with size. Fortunately, as can be seen from
Fig.~\ref{compar_prob}, for moderate values of $N$ the
distributions are not very different, and good results were
obtained in Refs.~\cite{koniar,deguchi} by this method.
Alternatively, Yao et al.\cite{yao} used a Monte Carlo pivoting
method\cite{mos} to generate many SAL configurations starting from
an initial loop on a cubic lattice. While this method produces
correctly weighted configurations, they may be statistically
dependent. Cognizant of this difficulty, Yao et al. sampled the
system at time intervals significantly exceeding the decay time of
geometrical correlations, and verified that the times of
appearance of knots behave like in a Poisson process.

In this work we employ a {\em direct unbiased} approach to
generating $(2N)$--step SALs. We first use successive
dimerizations to generate two $N$--step SAWs, and then check
whether (if starting from the same origin) they form a SAL. As
discussed before,  the probability of successful pairing of two
SAWs decays as $\sim N^{-2.08}$; this includes the rejection
probability that two segments of a formed loop intersect. We find
that the latter probability has only a weak dependence on the size
of the loop. Indeed, the distribution of the distance between the
origin and the 128th monomer in 256--step SALs, depicted by full
circles in Fig.~\ref{compar_prob}, practically coincides with the
probability distribution (dashed line) of generating loops from
two SAW segments, irrespective of their mutual intersections.

\begin{figure}
\includegraphics[width=8cm]{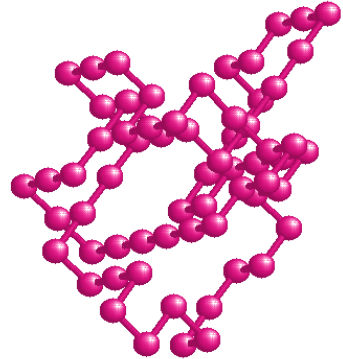}
\caption{\label{knot64} A trefoil knot in a 64--step SAL on a cubic lattice.}
\end{figure}

The small probability of two SAWs forming a loop can be enhanced
48-fold, by taking advantage of symmetries of the lattice to
consider only SAWs whose end--point coordinates $\vec{r}=(x,y,z)$
satisfy the relations $x\ge y \ge z\ge 0$. This is achieved by
generating a regular $N$-step SAW, and then performing the
following transformations: If the coordinate $x$ of the end--point
is negative, the walk is reflected with respect to $y-z$ plane;
and similarly for the other end--point coordinates. If the
end--point has $x<y$ then   $x$ and $y$ coordinates are
interchanged along the whole walk; with corresponding
interchanges for other pairs of axes. As a result of these
transformations, we reduce the space in which the end--point can
be located by a factor of 48, and thus increase the probability of
forming a loop. In continuum, such reduction of space does not
introduce any bias, since each SAW in the allowed subspace
corresponds to 48 SAWs in the original space. This is, however,
not true for SAWs ending on the boundaries of the allowed
subspace; e.g. there are only 24 SAWs with $x=y>z>0$, and
consequently loops created by SAWs ending at such points are
under--represented by a factor of 2. In continuum, configurations
with $x=y$ have zero measure, but on discrete lattices a
finite fraction of loops have this property, thus, creating a
slightly biased sample. Since this is a boundary effect, it
decreases with the inverse linear size of the space considered,
i.e. it will be proportional to $1/R$.

\begin{figure}
\includegraphics[width=8cm]{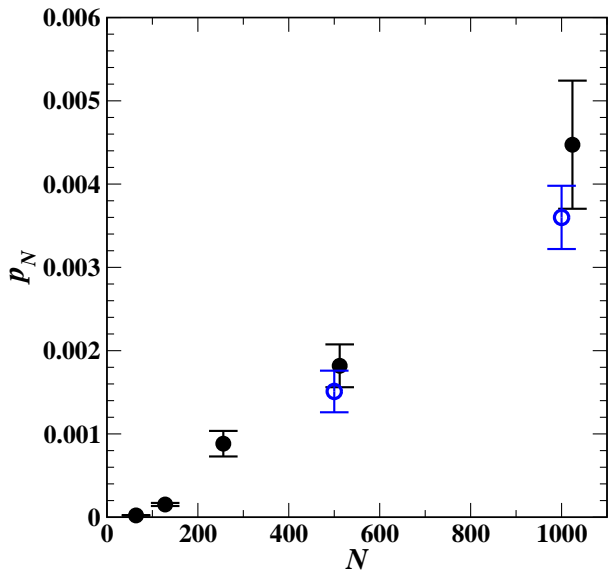}
\caption{\label{knot_prob} Full circles depict the probability
that an $N$-step SAL forms a knot. Error bars indicate one
standard deviation. For comparison, open circles show the results
of Ref.~\cite{yao}.}
\end{figure}

The direct method is quite efficient for generating small loops.
The minimal (trefoil) on a cubic lattice consists of 24
steps\cite{diao}, and the probability of its occurrence is
extremely small. For $N=64$, we had to examine 510 million pairs
of 32--step SAWs in order to find 1.4 million SALs, out of which
27 formed a trefoil knot of the type depicted in
Fig.~\ref{knot64}. At this point the probability to have a knot is
$0.000022\pm 0.000004$. We used the Alexander
polynomial\cite{alexander,alextheor} (at the value of its argument
equal to -1) to determine the presence and type of a knot. Since
almost all observed knots were trefoils, and very few more
complicated knots were encountered, this invariant sufficed for
our purposes. The limiting factor in our simulations was computer
time, and we could not go beyond $N=1024$, for which several
months of CPU on a multiple processor computer were necessary to
reach sufficient accuracy. At this value of $N$, the probability
to form SALs dropped to $10^{-5}$, even after the 48--fold
enhancement of the sampling. While the correction to bias in
sampling symmetric configurations was important for $N=64$, it
became negligible for $N=1024$.

Figure~\ref{knot_prob} depicts our results for the probability $p_N$
of having a knot in a SAL of length $N$. The curvature in the
results for this range of values of $N$ prevents a reliable
extraction of $N_o$ by a linear fit. Our two highest data points
compare well with the lowest data points of Yao et al.\cite{yao}
that are included for comparison. While our results give a
slightly larger probability than in Ref.~\cite{yao}, the
difference is only one standard deviation, and is not significant
at this level of accuracy. We also note that out of the 59 knots
detected for $N=512$, only two were not trefoils, consistent with
the results of Ref.~\cite{yao}.

In conclusion, we used the most direct method for generating an
unbiased ensemble of SALs on a cubic lattice (up to symmetry
factors). The method does not require large memory, but is very
time consuming. While quite efficient for small and moderate sized
SALs, it is limited to loops of around 1000 steps, at which the
probability of forming a knot is around half a percent. Our
results are consistent with those of Ref.~\cite{yao}, and provide
an independent support of the conclusions from Monte Carlo
sampling.

\acknowledgments
This work was supported by US-Israel Binational
Science Foundation (grant 1999-007), and  by the National Science
Foundation (grant DMR-01-18213).

\end{document}